# Optimal Control of Applications for Hybrid Cloud Services

Evgeniy Pluzhnik, Evgeniy Nikulchev, Simon Payain
Dep. of Cloud and Network Tech.,
Moscow Technological Institute
Moscow, Russia
e-mail: e.pluzhnik@gmail.com, e_nikulchev@mti.edu.ru, sadsema@gmail.com

*Abstract* - **Development of cloud computing enables to move Big Data in the hybrid cloud services. This requires research of all processing systems and data structures for provide QoS. Due to the fact that there are many bottlenecks requires monitoring and control system when performing a query. The models and optimization criteria for the design of systems in a hybrid cloud infrastructures are created. In this article suggested approaches and the results of this build.**

## I. INTRODUCTION

At the moment the Big Data issues develop. Fairly natural to query processing in cloud technologies, namely hybrid cloud infrastructure. However, most of the problems are not resolved, it is economic and technical and communication issues. It is important to develop technologies for porting applications lossless data access speed and providing QoS.

To optimize queries traditionally used model graph theory, algorithm theory and other methods of discrete mathematics. They cannot get the correct assessment of conformity of theoretical research into practical implementations in the cloud technologies. This is largely due to the fundamentally different building information systems it is not only the distributed data warehouse, but also the use of virtualization with dynamic reallocation of resources, the use of communication channels with different bandwidth for query processing, various platform features. All this has led to the search for new tools description of database management systems.

There are necessary to consider dynamic character of data exchange. Because communication channels, resource constraints, each block any application in the cloud must be evaluated during the execution at each stage. The dynamic descriptions in the form of approximation of differential equations with the vector control were used in the different tasks that are close to the considered in this.

In the first part of the article sets out the basic problem, based on the introduction of feedback. The second shows the models and criteria. In the third section, the algorithm control QoS. The fourth shows the results of processing queries to the data transferred to the hybrid cloud.

## II. OBJECTIVES AND TASKS

In [1] defined the basic principles of automated control systems for cloud computing resources. The main components of the architecture are:
- The use of SaaS-portal – allows for personalization of access to applications;
- Autonomous system of automatic control and Framework – includes implementation of the principles of autonomous control, including optimization modules [2], while applications and QoS-planner added as custom plugins;
- A hybrid cloud IaaS – integrated use of private cloud and public cloud services [3].

For definiteness, we consider scientific and educational resources of the educational institution. At the moment it is multimedia and video, and large amounts of text and other information. Moscow Institute of Technology is the largest Russian university, conducting distance learning and having a powerful data structure.

To form a block of workflow management system to query scientific and educational content of the experimental studies, this can be schematically represented as in Figure 1.

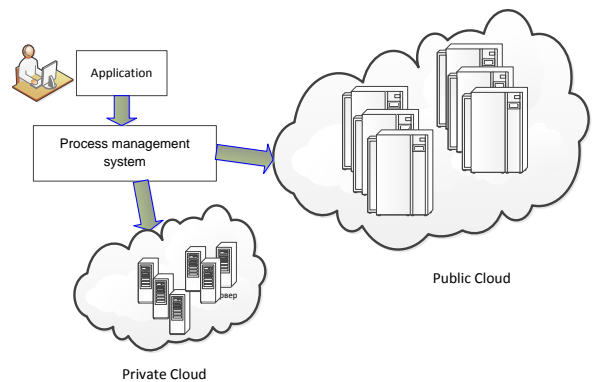

Figure 1.  The structure of the control system requests

To build information systems research and educational content with semistructured data necessary to solve the following problems:

*1)* Evaluation of the general parameters of the system (maximum number of users for simultaneous operation , the possibility of scaling services, the availability of personalized access).

*2)* Project cost estimate (having our own server capacity, cost comparison with the cost of building rental services).

*3)* Evaluation time data access, query performance evaluation for cloud infrastructures.

*4)* Construction of automatic allocation system and send requests in a distributed database.

### III. MODEL AND CRITERIA

To build dynamic models of governance must enter feedback. There are presences of feedback, first defined by the need to consider the current state of the virtual machine workload. Secondly for distributed databases final data are delivered to the client via the cloud environment, and information about the end of the query to be delivered to the central system. It is understood that the presence of more feedbacks charged already narrow communication channels and may increase the processing time. This is one more argument in favor of the control systems for the optimization of feedback. Feature of the cloud is positive feedback.

The base model offers classic model is approximated by the following difference equations,

$$x(t+1) = Ax(t) + Bu(t),$$

where $x = (x_1, x_2, ..., x_n)^T$ – $n$-dimensional vector of the system states under given constraints $x \in X \subseteq R^n$, $u = (u_1, u_2, ..., u_m)^T$ – $m$-dimensional vector of controls under given constraints $u \in U \subseteq R^m$, $t$ – discrete time instant.

In previous articles on the use of the control CPU, control of power in the data center and cloud computing [4, 5], a linear stationary system implementation found to be adequate.

Here the matrix is defined by *A* is determined from the structural relations system. For different variants of the decomposition of the local system when moving it to the cloud is your view matrix. Criterion transfer efficiency will increase query performance.

Obviously, this criterion cannot be specified explicitly in the proposed model form. Another important criterion will be minimizing management costs.

Can also be added to the criteria, taking into account the structural complexity and decomposition conditions – providing stability and control [6, 7].

However, given that the network is the bottleneck link exchange in a hybrid cloud infrastructure, increase the speed of the mechanism is to manage QoS.

### IV. MECHANISMS TO ENSURE QUALITY CRITERIA

Consider the standard network structure with hybrid cloud (Fig. 2). There shows a basic map of virtualization technology used in cloud infrastructure. There are three levels of allocated. At the first level two zones - local server, server for virtualization. Second level is the network which occurs switching, routing and data transmit. The third level is the public cloud and the global network.

Private cloud – level of hardware virtualization with support for special processor architecture. Local server level is a set of servers without of virtualization, where each server has a single operating system. Exchange of data between servers occurs at the level of switching and routing. At the Network level occurs packet routing, setting their priorities to control the queues, service levels and capacity allocation between the transmitted data types.

At the level of Public cloud computing is an allocation of the necessary resources for processing and storage of information without the possibility of changes in their allocation and distribution.

The transmission of information at the level Network and its preparation for transfer to the Internet, requires a dynamic control and management to improve the performance of applications in hybrid clouds.

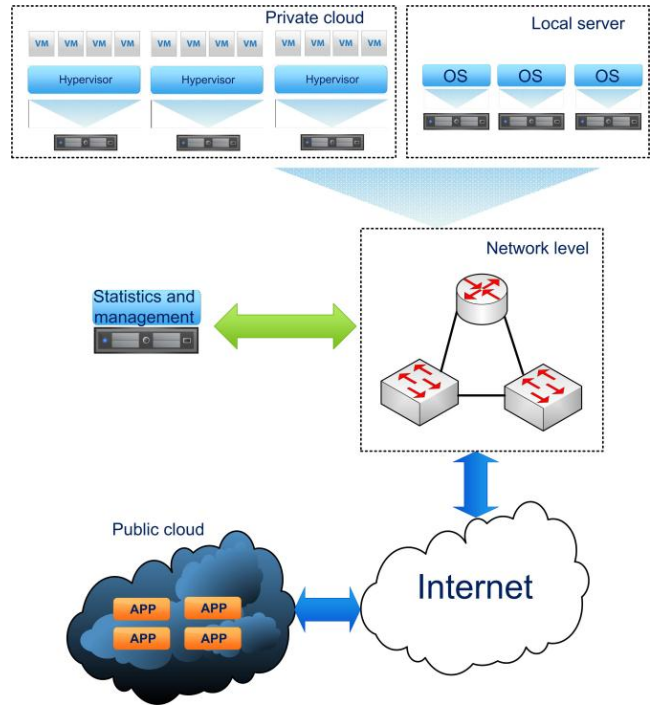

Figure 2. Network structure

Center of the routing is built on hardware cisco, using technology Cisco Application Visibility and Control. Cisco AVC solution is a services in network devices that provides application classification, monitoring, and traffic management to improve application performance with different quality of service (QoS) [8].

Separation of traffic types for monitoring and control is divided into two stages.

1. Traffic separation based on membership ip addresses and ports network (using access lists and policies):

A – private cloud and local servers traffic
B – public cloud traffic
C – other

2. Traffic types A and B separated by type of used applications using protocols NBAR2 and Flexible Netflow, which gives information about the use of bandwidth activity. Using Performance Agent data are obtained on transaction time, and the response delay of computing resources and other performance metrics by application (division by groups application)

a – web application
b - database application and requests
c – file application
d – service application traffic for the operation of systems

Based on the data traffic characteristics using the proposed scheduling algorithm is performed by each type of traffic management using Cisco AVC:

- Dynamically set different QoS priorities for applications type;
- Dynamically allocate bandwidth.

Based on data on the Internet download channel data obtained during the monitoring of the corporate network for each month, measured throughout the year, was built empirical histogram frequency channel load (Fig. 3a). Statistics obtained by information from the router interfaces on the amount of data transferred and loading port.

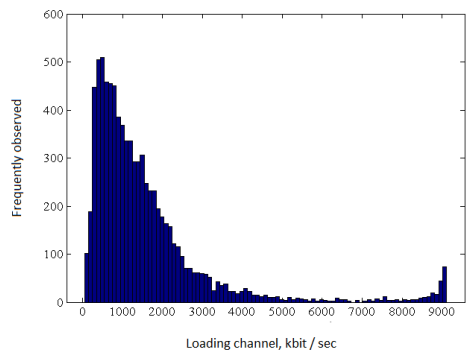

a)

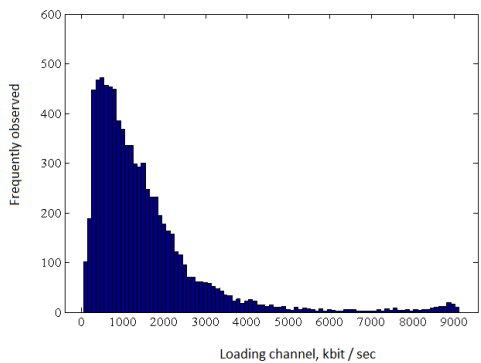

b)

Figure 3. Results of studies: a) without control, b) with control.

As a result of the control algorithm developed for the picture has changed, the histogram shown in Fig. 4b. It is seen that lost their tails in the distribution and peaks. This means that the number of dropped packets dropped and increased quality of data delivery.

## V. DATABASE TO HYBRID CLOUD

The experiment represents the production of databases in a hybrid cloud based on a complex search query. The results were compared with the requests of the local client-server system with the same subject databases.

In the table "Articles" are stored articles, article size from 100 KB to 3 MB. In the table "Authers" provides data to the author. Table "AuthorOfArticles" website links with the article. One article may be one main author and several co-authors. In the test load can be from 0 to 9 coauthors. The article shall be in the format docx. Size articles: a maximum – 3 MB, Minimum - 0.1MB.

In a local database (articlesLocal) data about authors and articles are stored in a relational database MS SQL Server (see Fig. 4). The occupied memory on the database server (articlesLocal) 26667,25 MB.

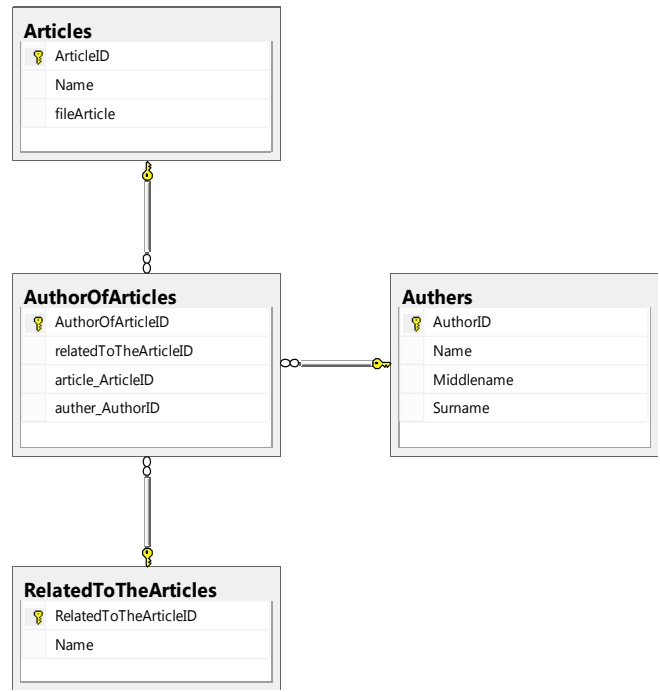

Figure 4. Database "articlesLocal"

The structure of the hybrid database is shown in Fig. 4.

In a hybrid DB (articlesHybrid) information about the authors and articles in the local database to MS SQL Server, and the article body - cloud storage Azure Storage. The occupied memory on the database server (articlesHybrid) 47,08 Mb, in the cloud, about 27 GB.

Interaction with the local database is organized by the following way:

1. The client application accesses the service to request articles for the parameter as a parameter was elected by the author of the article.

2. Service calls to the database to SQL Server.

3. SQL returns metadata and text articles service.

4. The service sends the result to the client request.

Interaction with the local database as follows.

Working with hybrid storage:

1. The client application accesses the service request articles of the parameter (a parameter was chosen author of the article).

2. Service calls to the database to SQL Server.

3. SQL returns service metadata of articles and addresses to which the texts are in Windows Azure.

4. The client receives from the service metadata of articles and addresses.

5. According to the obtained addresses the client accesses the cloud storage and receives text articles.

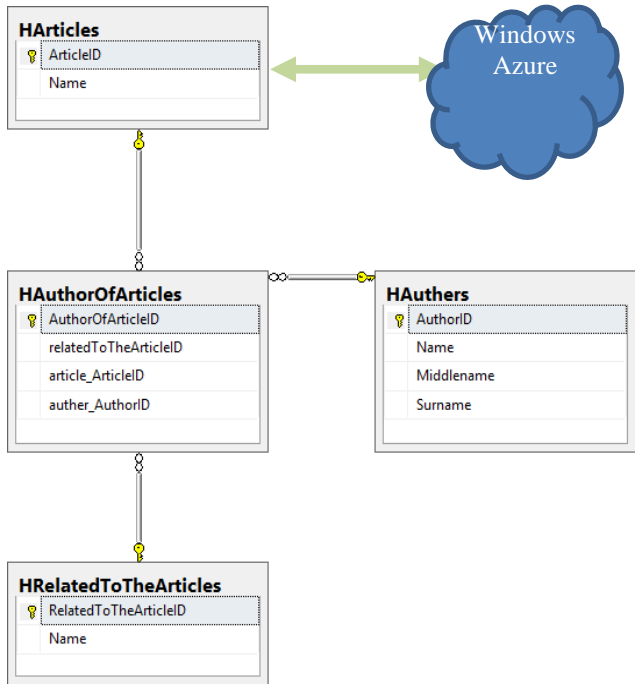

Figure 5. Database "articlesHybrid"

Results Tests were carried out queries to databases on mining of articles. The results of the first experiment are shown in Figure 6.

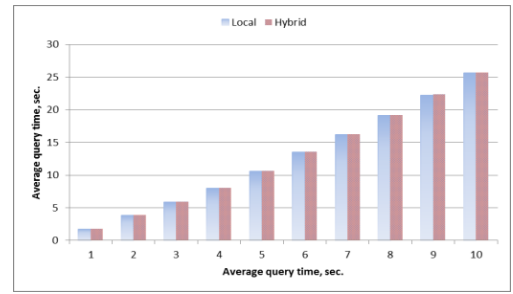

Figure 6. Results of test.

The experiment demonstrates the effectiveness of cloud storage for systems using semistructured database.

Based on the results we can conclude that, for the time evaluation of database queries application of graph theory will not be effective. Resources used in the processing of queries on the local server structure other than in an external cloud storage.

REFERENCES

[1] S. Babu, J. Chase and S. Parekh, "Automated control in cloud computing: challenges and opportunities", Proc. 1st workshop on Automated control for datacenters and clouds, 2009, pp. 13–18, doi: 10.1145/1555271.1555275

[2] S. Haak and M. Menzel, "Autonomic benchmarking for cloud infrastructures: an economic optimization model" Proc. ACM/IEEE Workshop on Autonomic Computing for Economics, 2011, pp. 27-32 doi: 10.1145/1998561.1998569

[3] E.V. Pluzhnik and E.V. Nikulchev, "Use of dynamical systems modeling to hybrid cloud database" Int'l J. of Communications, Network and System Sciences, vol. 6. no. 12, 2013, pp. 505-512, doi: 10.4236/ijcns.2013.612054

[4] R. Nathuji, A. Kansal and A. Ghaffarkhah, "Q-clouds: Managing performance interference effects for QoS-aware clouds", Proc. the ACM European Society in Systems Conference. Paris, France, 2010, pp. 237–250, doi: 10.1145/1755913.1755938

[5] J. M. Luna and C. T. Abdallah, "Control in computing systems: Part I" Proc. IEEE International Symp. Computer-Aided Control System Design (CACSD), 2011, pp. 25-31, doi: 10.1109/CACSD.2011.6044541)

[6] M. D. Lemmon "Towards a passivity framework for power control and response time management in cloud computing" Proc. of 7th Intl. Workshop on Feedback Computing, San Jose, CA. 2012 (http://controlofsystems.org/FeedbackComputing2012/papers/Sess13-feedbackcomputing2012.pdf)

[7] E.V. Pluzhnik, E.V. Nikulchev and S.V.Payain "Concept of Feedback in Future Computing Models to Cloud Systems" Proc. Conf. ISC'14, in press

[8] T. Bujlow and V. Carela-Espanol, "Comparison of Deep Packet Inspection (DPI) Tools for Traffic Classification", Technical Report, UPC, 2013, http://vbn.aau.dk/ws/files/78068418/report.pdf